\documentstyle[emulateapj,onecolfloat,psfig]{article}
\begin{document}
\twocolumn[
\title{47~Tuc: The Spectroscopic Versus CMD Age Discrepancy}

\author{Alexandre Vazdekis \altaffilmark{1}}
\affil{$^1$ Deptartment of Physics, University of Durham, Durham DH1 3LE, U.K.\\
E-mail: Alexandre.Vazdekis@durham.ac.uk}

\author{Maurizio Salaris \altaffilmark{2}}
\affil{$^2$ Liverpool John Moores University, Twelve Quays House, Egerton Wharf, Birkenhead
CH41 1LD, U.K.\\
E-mail: ms@astro.livjm.ac.uk}

\author{Nobuo Arimoto \altaffilmark{3}}
\affil{$^{3}$ Institute of Astronomy, University of Tokyo, Osawa 2-21-1, Mitaka, Tokyo
181-0015, Japan\\
E-mail: arimoto@mtk.ioa.s.u-tokyo.ac.jp}

\and

\author{James A. Rose \altaffilmark{4}}
\affil{$^4$ Department of Physics and Astronomy, University of North Carolina, Chapel Hill, NC
27599, U.S.A.\\
E-mail: jim@physics.unc.edu}

\submitted{Accepted for publication in The Astrophysical Journal, vol. 549, Mar 1, 2001 issue}

%%%%%%%%%%%%%%%%%%%%%%%%%%%%%%%%%%%%%%%%%%%%%%%%%%%%%%%%%%%%%%%%%%%%%%%%%%%%%%%%%%%%%
%%%%%%%%%%%%%%%%%%%%%%%%%%%%%%%%%%%%%%%%%%%%%%%%%%%%%%%%%%%%%%%%%%%%%%%%%%%%%%%%%%%%%
\begin{abstract}
%%%%%%%%%%%%%%%%%%%%%%%%%%%%%%%%%%%%%%%%%%%%%%%%%%%%%%%%%%%%%%%%%%%%%%%%%%%%%%%%%%%%%
%%%%%%%%%%%%%%%%%%%%%%%%%%%%%%%%%%%%%%%%%%%%%%%%%%%%%%%%%%%%%%%%%%%%%%%%%%%%%%%%%%%%%
We investigate current problems in obtaining reliable ages for old stellar
systems based on stellar population synthesis modelling of their integrated
spectra.  In particular, we address the large ages derived for the globular
cluster 47~Tuc, which is at odds with its Color-Magnitude-Diagram (CMD)
age.  Using a new age indicator, H$\gamma_{\sigma<130}$, which is
particularly effective at breaking the degeneracy between age and metallicity,
we confirm the discrepancy between the spectroscopic age and the CMD age of
47~Tuc, in that the spectroscopic age is much older.  
Nebular emission appears unlikely to be a source for weakening the observed
Balmer lines. We then explore a number of key parameters
affecting the temperature of Turn-Off stars, which are the main contributors
to the Balmer lines for old metal-rich stellar populations.  We find that
$\alpha$-enhanced isochrones with atomic diffusion included not only provides
a good fit to the CMD of 47~Tuc, but also leads to a spectroscopic age in
better agreement with the CMD age.

\end{abstract}
]

%%%%%%%%%%%%%%%%%%%%%%%%%%%%%%%%%%%%%%%%%%%%%%%%%%%%%%%%%%%%%%%%%%%%%%%%%%%%%%%%%%%%%
%%%%%%%%%%%%%%%%%%%%%%%%%%%%%%%%%%%%%%%%%%%%%%%%%%%%%%%%%%%%%%%%%%%%%%%%%%%%%%%%%%%%%
\section{Introduction}
%%%%%%%%%%%%%%%%%%%%%%%%%%%%%%%%%%%%%%%%%%%%%%%%%%%%%%%%%%%%%%%%%%%%%%%%%%%%%%%%%%%%%
%%%%%%%%%%%%%%%%%%%%%%%%%%%%%%%%%%%%%%%%%%%%%%%%%%%%%%%%%%%%%%%%%%%%%%%%%%%%%%%%%%%%%

An estimate of the mean luminosity-weighted stellar age of an early-type galaxy 
represents a major step in unveiling its true star formation history. However, 
to derive reliable information about stellar ages from the integrated light of 
unresolved galaxies one must deal with the age-metallicity degeneracy problem, 
which affects not only  
integrated colours but also absorption line-strengths (Worthey 1994). 
Recently, new age-dating techniques based on the Balmer lines (e.g.,
Jones \& Worthey 1995; Vazdekis \& Arimoto 1999, thereafter VA99) have shown 
great promise in untangling the age-metallicity degeneracy.  These techniques
should be tested and calibrated on the metal-rich Galactic globular clusters 
(GCs) for which, unlike elliptical galaxies, independent age estimates are 
possible (Gibson et al. 1999, thereafter G99) by means of the 
Color-Magnitude-Diagram (CMD) of their resolved stellar population.
 
The application of the new age-dating techniques to very high signal-to-noise 
ratio (S/N) spectra of metal-rich Galactic GCs has revealed two major concerns: 
{\it i)} the obtained ages are unreasonably large ($>$20~Gyr) (Jones 1999; 
Cohen, Blakeslee \& Ryzhov 1998; G99; VA99) and {\it ii)} a severe disagreement 
is found between the spectroscopic and CMD ages of 47~Tuc (G99). 
The CMD-derived ages may be sensitive to the dating method employed (e.g.,
Alonso et al. 1997) and to the input physics of the theoretical isochrones used 
as a reference (e.g., Salaris \& Weiss 1998, thereafter SW98). However, a
variety of recent CMD-based age determinations for 47~Tuc have consistently
found its age to lie within 9-12.5 Gyr (SW98; Gratton et al.~1997;
Carretta et al.~2000; Liu \& Chaboyer~2000).

These CMD-derived ages are substantially younger than the $>20$~Gyr 
spectroscopically derived age obtained by G99 using the age indicator
of Jones \& Worthey (1995), and Worthey (1994) stellar population 
models. This discrepancy shows clearly that current stellar population 
synthesis models used for interpreting the integrated light of  
stellar systems may have severe zero points problems. 

In \S~2 we improve the spectroscopic age-dating technique which confirms a 
large spectroscopic age for 47~Tuc. In \S~3 we discuss the possible
origin of the problem, explore several theoretical parameters and
suggest a possible solution. Finally, in \S~4 we present our conclusions.

%%%%%%%%%%%%%%%%%%%%%%%%%%%%%%%%%%%%%%%%%%%%%%%%%%%%%%%%%%%%%%%%%%%%%%%%%%%%%%%%%%%%%
%%%%%%%%%%%%%%%%%%%%%%%%%%%%%%%%%%%%%%%%%%%%%%%%%%%%%%%%%%%%%%%%%%%%%%%%%%%%%%%%%%%%%
\section{A new H$\gamma$ age indicator: 47~Tuc age estimate}
%%%%%%%%%%%%%%%%%%%%%%%%%%%%%%%%%%%%%%%%%%%%%%%%%%%%%%%%%%%%%%%%%%%%%%%%%%%%%%%%%%%%%
%%%%%%%%%%%%%%%%%%%%%%%%%%%%%%%%%%%%%%%%%%%%%%%%%%%%%%%%%%%%%%%%%%%%%%%%%%%%%%%%%%%%%
In this study we make use of the evolutionary stellar population synthesis 
model of Vazdekis (1999) (hereafter V99), which predicts spectral energy 
distributions in the optical wavelength range for single burst old-aged 
stellar populations of metallicities ($-0.7\leq\log(Z/Z_\odot)\leq+0.2$), at 
resolution 1.8\AA~(FWHM). This approach is different from the one followed
by previous models (e.g., Worthey 1994; Vazdekis et al. 1996) which
used mostly the Lick/IDS polynomial fitting functions (Worthey et al. 1994; 
Worthey \& Ottaviani 1997) to relate 
the strengths of selected absorption features to stellar atmospheric 
parameters. These fitting functions are based on
the Lick/IDS stellar library (FWHM$\sim$9~\AA, Worthey et al. 1994), 
thus limiting the sensitivity of the models to weak features.
However V99 model provides full SEDs rather than predicted index 
strengths and therefore the Lick indices defined in Worthey et al. (1994)
and Worthey \& Ottaviani (1997) as well as those of Rose (1994) and
Jones \& Worthey (1995) are measured directly on the SEDs of V99 models
(without the use of any fitting functions). This approach makes it easy
to define new indices and even to confront the 
model predictions to the detailed structure of observed absorption features.
V99 models have been recently updated 
(Vazdekis 2000, in preparation) with Girardi et al. (2000) scaled-solar 
isochrones and new empirical photometric libraries, such as 
Alonso et al. (1999).

The chief impediment to obtaining reliable ages of elliptical galaxies is 
caused by the degenerate effects of age and metallicity on the integrated 
spectra of old stellar populations (e.g., Worthey 1994).  Recently, VA99
have proposed a new age indicator, centered on H$\gamma$, which provides 
unprecedented power for breaking the age-metallicity degeneracy.  
Their index is a pseudoequivalent width measurement, relying on the 
pseudocontinuum peaks immediately longward and shortward of H$\gamma$, and 
is an improvement over earlier pseudoequivalent width H$\gamma$ indices 
(Rose 1994; Jones \& Worthey 1995) in being relatively insensitive to spectral 
resolution. We have redefined the VA99 index, to take full 
advantage of the information available in an integrated spectrum with 
$\sigma\sim100~km s^{-1}$ and to make the index as insensitive as possible to 
metallicity. This new index, H$\gamma_{\sigma<130}$, is composed of two 
pseudocontinua, $\lambda\lambda$ 4329.000-4340.468\AA\ and 4352.500-4368.250\AA, 
and the feature at $\lambda\lambda$ 4333.250-4363.000\AA. The main 
difference with respect to the VA99 definition is that the feature now also
covers the neighboring metallic line centered on $\lambda\sim$4352\AA.
The two pseudocontinua severely overlap the index passband aiming at making
H$\gamma_{\sigma<130}$ stable against changes in the spectral resolution (see below),
and insensitive to metallicity variations on the basis of the compensating 
effect raised up by VA99: at a given age, H$\gamma$ strengthens with 
metallicity owing to the adjacent metallic absorption, but on the other
hand the pseudocontinua are depressed by the effects of the neighboring Fe{\sc I}
lines on both sides of H$\gamma$ (see VA99 for an extensive explanation).

\begin{figure}[t]
\centerline{\psfig{file=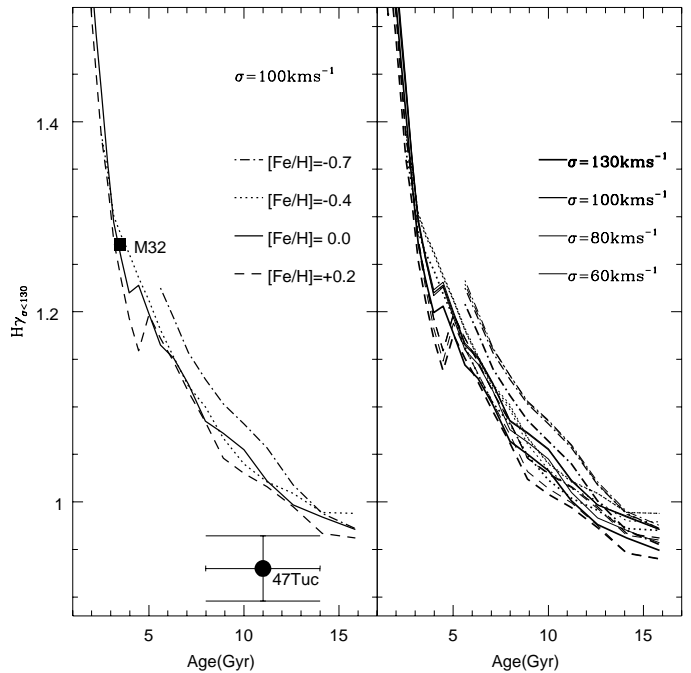,width=3.8in}}
\vskip-0.3cm
\caption{
Left panel: the new H$\gamma_{\sigma<130}$ age indicator measured on V99 model spectral 
library smoothed to $\sigma\sim100~km s^{-1}$ to match the resolution of the very high S/N 
spectrum of 47~Tuc (Rose 1994). The size of the error bar along the age axis represents 
the maximum range for the most recent CMD age estimates ($9\pm1$~Gyr in SW98 and 
$12.5\pm1.5$~Gyr in Liu \& Chaboyer~2000), and therefore we plotted 47~Tuc value at 11~Gyr. 
Right panel: test for the stability of H$\gamma_{\sigma<130}$ against resolution in the 
range $60 < \sigma < 130~km s^{-1}$. 
}
\label{fig:Hg130}
\end{figure}

Fig.~\ref{fig:Hg130} shows the age disentangling power of the H$\gamma_{\sigma<130}$. 
In the left panel, the indices for both M~32 (Jones 1999 spectrum) and for 
47~Tuc (Rose 1994 spectrum) are plotted relative to the models. The age estimate 
for M~32 is consistent with VA99 result ($\sim$4~Gyr), based on the earlier 
H$\gamma$ index.  The principal issue for this paper, however, {\it is the 
very large age inferred for 47~Tuc, i.e., well in excess of 15~Gyr}. This 
disturbingly large inferred age confirms that already obtained by G99 and VA99, 
and thus is a feature of all recent studies of this cluster (based on different
observational spectra). The right panel 
of Fig.~\ref{fig:Hg130} shows the insensitivity of H$\gamma_{\sigma<130}$ to 
resolution in the range $60 < \sigma < 130~km s^{-1}$. Spectra of 
S/N(\AA)$\sim$175 and a very careful correction of any $\lambda$ shift 
(see VA99), are required to take the full advantage of this index. 

\begin{figure}[t]
\centerline{\psfig{file=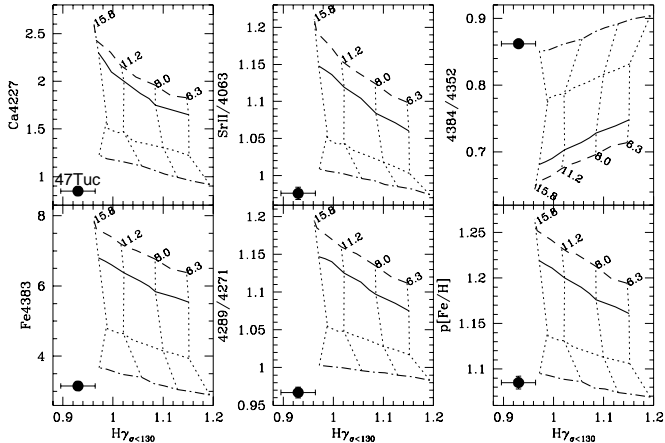,width=3.75in}}
\vskip-3.1cm
\caption{
H$\gamma_{\sigma<130}$ versus different indices as defined in Worthey et al. (1994) 
(Ca4227; Fe4383) and Rose (1994). All these indices were measured on the two, 
the 47~Tuc spectrum of Rose (1994), and V99 models smoothed to $\sigma=100~kms^{-1}$. 
Line types as in Fig.~\ref{fig:Hg130}, while thin dotted lines mean models of equal ages indicated 
by the numbers (Gyr).
}
\label{fig:indices}
\end{figure}

Fig.~\ref{fig:indices} further illustrates the age-metallicity resolving power achieved 
with the H$\gamma_{\sigma<130}$ index, and further clarifies the troublesome problem 
of the inferred age for 47~Tuc. Here, H$\gamma_{\sigma<130}$ is plotted versus 
several different indices defined in Worthey et al.~(1994) and Rose (1994). All the
indices were measured directly on the model SEDs of V99 smoothed to match
the resolution of the spectrum of 47~Tuc.  
Again, all index plots indicate an age for 47~Tuc in excess of 15 Gyr. 
In addition, most of the plots in Fig.~\ref{fig:indices} suggest a metallicity for the 
cluster around 0.1-0.2 dex lower than the Carretta \& Gratton (1997) value of 
[Fe/H]=$-$0.7, which is based on high dispersion spectra of individual red 
giants.  On the other hand, Gratton \& Sneden (1991) and 
Brown \& Wallerstein (1992), obtained [Fe/H]$\sim-$0.8 and $-$0.9, 
respectively. Recently, Liu \& Chaboyer (2000) obtained [Fe/H]=-0.95 using the 
isochrone fitting technique.  However, it bears emphasizing that the uncertainty
in metallicity does not affect the age discrepancy shown here.

%%%%%%%%%%%%%%%%%%%%%%%%%%%%%%%%%%%%%%%%%%%%%%%%%%%%%%%%%%%%%%%%%%%%%%%%%%%%%%%%%%%%%
%%%%%%%%%%%%%%%%%%%%%%%%%%%%%%%%%%%%%%%%%%%%%%%%%%%%%%%%%%%%%%%%%%%%%%%%%%%%%%%%%%%%%
%%%%%%%%%%%%%%%%%%%%%%%%%%%%%%%%%%%%%%%%%%%%%%%%%%%%%%%%%%%%%%%%%%%%%%%%%%%%%%%%%%%%%

%%%%%%%%%%%%%%%%%%%%%%%%%%%%%%%%%%%%%%%%%%%%%%%%%%%%%%%%%%%%%%%%%%%%%%%%%%%%%%%%%%%%%
%%%%%%%%%%%%%%%%%%%%%%%%%%%%%%%%%%%%%%%%%%%%%%%%%%%%%%%%%%%%%%%%%%%%%%%%%%%%%%%%%%%%%
\section{Discussion}
%%%%%%%%%%%%%%%%%%%%%%%%%%%%%%%%%%%%%%%%%%%%%%%%%%%%%%%%%%%%%%%%%%%%%%%%%%%%%%%%%%%%%
%%%%%%%%%%%%%%%%%%%%%%%%%%%%%%%%%%%%%%%%%%%%%%%%%%%%%%%%%%%%%%%%%%%%%%%%%%%%%%%%%%%%%
In this section we explore possible causes for the 47 Tuc age discrepancy: emission
line contamination, possible peculiarity of the cluster, horizontal branch
contribution, mixing length calibration, $\alpha$-elements enhancement, initial 
helium abundance and atomic diffusion.

\subsection{Nebular emission}
\begin{figure}[t]
\centerline{\psfig{file=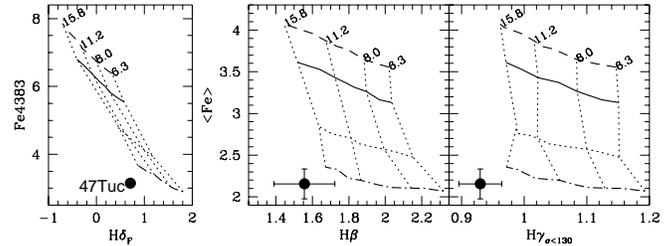,width=3.75in}}
\vskip-5.8cm
\caption{
Left: H$\delta_F$ as defined in Worthey \& Ottaviani (1997) is plotted 
versus Fe4383. 
Middle: H$\beta$ versus $<Fe>$ (i.e., $\frac{1}{2}(Fe5270+Fe5335)$) 
(index definitions from Worthey et al~1994). Since H$\beta$ is not covered 
by the spectral range of the spectrum of Rose (1994) we used the spectrum of 
Covino et al (1995) with FWHM=3.4\AA. 
V99 models were smoothed to match this resolution. 
Right: H$\gamma_{\sigma<130}$ versus $<Fe>$ (H$\gamma_{\sigma<130}$ 
measured on the Rose 1994 spectrum). 
}
\label{fig:emision}
\end{figure}

G99 pointed out that gas emission could be partially filling 
in H$\gamma$. We therefore studied other Balmer lines, since emission would affect 
them by differing amounts, according to standard Case B recombination line 
physics (e.g., Osterbrock 1989). Fig.~\ref{fig:emision} shows that the derived age 
for 47~Tuc is essentially independent of the Balmer line used. If emission 
fill-in were the cause of the large spectroscopic 
age, one would expect the ages derived from H$\beta$ and H$\delta$ to be older 
and younger, respectively, than that obtained from H$\gamma$.  To quantify this
statement, we have used a flux-calibrated spectrum of the Orion Nebula as a
template for a Case B hydrogen recombination spectrum.  In fact, we adjusted
the Balmer line intensities to exactly match the Case B prescription.  We
then normalized the Orion spectrum so that when subtracted from the 47~Tuc
spectrum at H$\beta$, the derived age from the H$\beta$ index is reduced to
11 Gyr.  The ages derived from the H$\gamma_{\sigma<130}$ and H$\delta_F$
indices are $\sim$12.5 Gyr and $\sim$14 Gyr.  Thus the emission-corrected
ages from different Balmer lines are, in fact, discordant, although the degree
of discrepancy is probably not sufficient to eliminate the emission hypothesis
altogether.

An additional problem for the emission hypothesis comes from the fact
that spectra of other metal-rich Galactic globular clusters
exhibit the same ``anomalous'' behavior in H$\gamma$ as 47~Tuc.  Specifically,
in Rose (1994) a spectrum formed from the composite of four metal-rich
globulars (NGC6356, NGC6624, M69, and M71) is compared to that of 47~Tuc (and
M32), and shown to be nearly indistinguishable with 47~Tuc in all respects,
including the pseudoequivalent width index for H$\gamma$.  We have further
quantified this fact by measuring the H$\gamma_{\sigma<130}$ indices for the
four clusters.  On average, H$\gamma_{\sigma<130}$ for the four clusters is
0.87, i.e., slightly lower even than for 47~Tuc.  Thus if the
emission hypothesis is correct, then in composite, the other four clusters
must have essentially the same amount of emission fill-in as 47~Tuc.  Such a
scenario appears to be rather contrived.  In short, emission fill-in appears
to be an unlikely source of the weak Balmer lines in 47~Tuc, but a more
definitive assessment of the emission hypothesis could be
achieved via deep interference filter imaging of the cluster.

In mentioning the other four metal-rich globular clusters it is also worth 
noting that there are still some integrated light properties of metal-rich clusters
(e.g., anti-correlation between strengths of CN bands and the SrII$\lambda$4077
line) that remain unexplained, both among Galactic (Rose \& Tripicco
1986) and M31 globular clusters (Burstein et al. 1984; Tripicco 1989).

\subsection{Is 47~Tuc a peculiar cluster?}
A second possibility is that 47~Tuc is peculiar among Galactic GC's. However,
by sketching the models of Vazdekis et al. (1996) to the Mg$_2$-H$\beta$ plot 
of Burstein et al. (1984) (c.f. their Fig.~5k) we see that a number of metal-rich
Milky Way GC's fall well below the model lines (e.g., models of [Fe/H]=-0.7
and 16~Gyr yield Mg$_2$$\sim$0.17 and H$\beta$$\sim$1.7). The same result is
achieved if Worthey (1994) model grids are used.
VA99, Vazdekis et al. (1996; c.f., their Fig.~8), and Cohen, Blakeslee \&
Ryzhov (1998) find analogous results for a set of metal-rich galactic GC's.
Finally, as mentioned above the composite spectrum of four metal-rich clusters
analyzed in Rose (1994) is similar in all respects to 47 Tuc.

%%%%%%%%%%%%%%%%%%%%%%%%%%%%%%%%%%%%%%%%%%%%%%%%%%%%%%%%%%%%%%%%%%%%%%%%%%%%%%%%%%%%%
%%%%%%%%%%%%%%%%%%%%%%%%%%%%%%%%%%%%%%%%%%%%%%%%%%%%%%%%%%%%%%%%%%%%%%%%%%%%%%%%%%%%%

\subsection{The Horizontal Branch contribution}
The Balmer indices are dominated by the hottest stars along the isochrone 
(e.g., Rose 1994; Worthey 1994; Buzzoni, Mantegazza \& Gariboldi 1994), i.e., 
by TO and Horizontal Branch (HB) stars. For a GC with a red HB such as 47~Tuc 
we have verified that the contribution due to HB stars is negligible for this 
index. In fact, a decrease of the HB stars temperature by 150~K yields 
H$\gamma_{\sigma<130}$ values smaller by less than 0.01\AA, in the range of 
ages 6-16~Gyr. Therefore,  the discrepancy between the integrated light and 
CMD ages must be found in the TO. On the basis of Fig.~\ref{fig:Hg130}
and the TO temperatures of our isochrones we estimate that to decrease 
H$\gamma_{\sigma<130}$ by $\sim0.050$\AA\ (to match the observed value for 
typical CMD ages) requires a TO cooler by $\sim$200~K, for a given age. We 
therefore should look at those  parameters affecting the TO effective temperature.

\subsection{Mixing length}
The stellar models we use adopt a solar calibrated value of the mixing
length. This assumption is
in agreement with results from current 2-D hydrodynamical simulation
of superadiabatic stellar convection, even for non solar metallicity
stars (Freytag \& Salaris 1999). However, we have also tested an alternative 
prescription for the treatment of superadiabatic convection in stellar 
envelopes, namely, the Full Spectrum Turbulence theory (FST -- see, e.g., 
D'Antona, Caloi \& Mazzitelli 1997). We find that for the age and metallicity
regime typical of 47~Tuc the 
difference of TO effective temperatures between solar calibrated mixing 
length and FST models are negligible ($\sim$10~K).

\subsection{$\alpha$-elements enhancement}
The chemical composition of 47~Tuc stars (and of Galactic GCs in general) 
is enhanced in $\alpha$ elements (we mean mainly O, Ne, Mg, Si, S, Ca, Ti -- 
see, e.g., the review by Carney 1996). Spectroscopic determinations of
$\alpha$ elements abundances in 47Tuc stars have been performed
by Gratton et al.~(1986), Brown et al.~(1990), 
Brown \& Wallerstein (1992), Norris \& Da Costa (1995); their results,
as summarized by Carney (1996), provide $<$[O/Fe]$>$=0.53$\pm$0.08
and an average enhancement of [Si+Ca+Ti/Fe] of about 0.20 dex.

Moreover, as demonstrated by SW98 and confirmed by 
VandenBerg, Swenson \& Alexander (2000), for the [$\alpha$/Fe] ratios 
observed in Galactic GCs ([$\alpha$/Fe]$\sim$0.3--0.4) and [Fe/H]$>$-1,
it is not valid to use scaled-solar isochrones with the same global 
metallicity as the $\alpha$-enhanced ones to approximate the effect of the 
$\alpha$-elements enhancement. As a preliminary test we computed 
selected scaled-solar 
isochrones using the same code and input physics as in SW98, providing
a good agreement for the TO temperatures with the corresponding 
isochrones of Girardi et al (2000) (used as scaled-solar ones in our 
population synthesis code).  
When comparing $\alpha$-enhanced with scaled-solar isochrones 
the former have a hotter TO if the comparison is made at the same
global metallicity (Z=0.008), while if the comparison is made at the
same [Fe/H]=$-$0.7, they have a cooler TO (see SW98). 

We therefore implemented in our population synthesis code 
the $\alpha$-enhanced isochrones ($< [\alpha/$Fe]$>$=0.4) of SW98, 
transformed to the observational plane following our empirical 
prescriptions (the same ones that we applied to the Girardi et al.~2000 
isochrones). The referee asked us to state that we did not change 
the input empirical stellar spectral database when calculating the 
new $\alpha$-enhanced SSP model SEDs and therefore the results shown 
here have the potential to be modified. 
Ideally, we should use stellar spectra of appropriate abundance
ratios, and this is exactly the case since our library of empirical
spectra is based on observations of local metal poor stars which show
the same abundance pattern (e.g., $\alpha$-elements enhancement) as in
galactic GCs (see, e.g., Table 1 in Salaris \& Weiss 1998 and
references therein). If there is any inconsistency between the
spectral library and the stellar models used to determine the age of 47Tuc
is when considering models with scaled solar abundances.
However, in this case  we selected the required 
stellar spectra following their [Fe/H] rather than their total metallicity Z. 
In this way, at least, the Fe lines estimates from the new model SEDs 
are more secure. On the other hand, the fact that the Balmer lines are 
being studied means that the opacity changes (according to the chosen 
abundance ratios) operate mostly in the pseudocontinua and may not have
a significative impact on the results discussed here.

\begin{figure}[t]
\centerline{\psfig{file=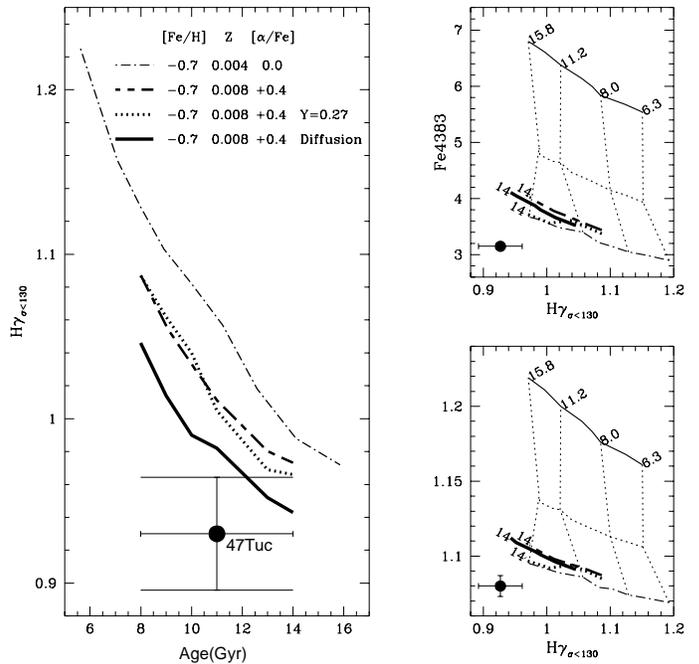,width=3.8in}}
\vskip-0.3cm
\caption{
Left: H$\gamma_{\sigma<130}$ measured on the integrated spectra synthesized on the 
basis of different isochrones. 
Right: varios metallicity indices are plotted versus H$\gamma_{\sigma<130}$. 
For the models represented by the thick lines the age vary from 8 to 14~Gyr. 
}
\label{fig:solut}
\end{figure}

Fig.~\ref{fig:solut} shows that the H$\gamma_{\sigma<130}$ values measured on 
the spectra synthesized on the basis of $\alpha$-enhanced isochrones yield 
considerably younger ages than the scaled-solar ones. To understand this 
point we recall that VA99 have clearly shown that for the higher spectral 
resolutions the weakening of the H$\gamma$ feature at $\lambda$ 4340\AA\ due 
to a higher metallicity (caused by the lower temperatures of TO stars due to 
higher opacity of the stellar matter) is compensated by a deepening of the 
adjacent iron lines (mainly at the blue side of H$\gamma$). The 
H$\gamma_{\sigma<130}$ index definition takes into account this effect, and 
thus is insensitive to metallicity variations. Using the $\alpha$-enhanced 
isochrones at a given [Fe/H] the global metallicity is higher and therefore 
the temperatures of the stars slightly lower, causing a weakening of the 
H$\gamma$ feature which is not compensated by any deepening of the adjacent 
iron lines. Therefore, this effect is very similar to the one produced by an 
age increase. Furthermore, we also have tested $\alpha$-enhanced isochrones 
with [Fe/H]=-0.3 (Z=0.02), confirming that for a given $\alpha$-enhancement 
the H$\gamma_{\sigma<130}$ index preserved its age resolving power.

%%%%%%%%%%%%%%%%%%%%%%%%%%%%%%%%%%%%%%%%%%%%%%%%%%%%%%%%%%%%%%%%%%%%%%%%%%%%%%%%%%%%%
%%%%%%%%%%%%%%%%%%%%%%%%%%%%%%%%%%%%%%%%%%%%%%%%%%%%%%%%%%%%%%%%%%%%%%%%%%%%%%%%%%%%%

\subsection{Primordial helium content}
We find that a reasonable variation of the initial He content does not 
appreciably affect the spectroscopic age. In Fig.~\ref{fig:solut} the 
H$\gamma_{\sigma<130}$ 
values obtained on the basis of $\alpha$-enhanced isochrones with a solar 
initial He abundance (Y=0.273) are not significantly different from the ones 
computed with Y=0.254 at the 47~Tuc metallicity ($\Delta$Y/$\Delta$Z=3).

\subsection{Atomic diffusion}
Atomic diffusion is capable of changing the TO temperature of low mass stars 
at a given age (see. e.g., Proffitt \& Vandenberg 1991, Chaboyer et al~1992, 
Castellani et al~1997, Cassisi et al~1998, Salaris, Groenewegen \& Weiss 2000). 
The occurrence of this physical process in the sun has been demonstrated by 
helioseismic studies (e.g., Guenther et al~1996). Lebreton et al. (1999)
showed that diffusion is required to reproduce the 
temperatures of Hipparcos subdwarfs with $-1.0<$[Fe/H]$<-0.3$. 
Due to diffusion, the surface metallicity and He content 
decrease during the Main Sequence phase (MS) due to their sinking below the
convective envelope. Around the TO the surface [Fe/H] and Y show a minumum; 
then evolutionary timescales become much shorter and diffusion is no longer 
effective. Moreover, since envelope convection deepens, almost all metals and 
He diffused toward the center are engulfed again in the convective envelope. 
Along the Red Giant Branch the surface [Fe/H] (and He) is 
restored to nearly its initial value (the current spectroscopic determinations 
of [Fe/H] for GCs make use of Red Giant stars, so that the measured metal 
abundances truly reflect the initial ones). For a given initial chemical 
composition and age, TO temperatures are significantly cooler than 
in isochrones without diffusion (the TO luminosities are reduced too, 
thus causing an age reduction by $\simeq$ 1 Gyr if the TO brightness is used 
as age indicator). The reason for this behavior is that the inward settling 
of He during the MS raises the core molecular weight and 
the molecular weight gradient between surface and center of the star. This 
increases the stellar radius and the rate of energy generation in the center. 
The diffusion of the metals partially counterbalances this effect by 
decreasing the opacity in the envelope and increasing the central CNO abundance.

\begin{figure}[t]
\centerline{\psfig{file=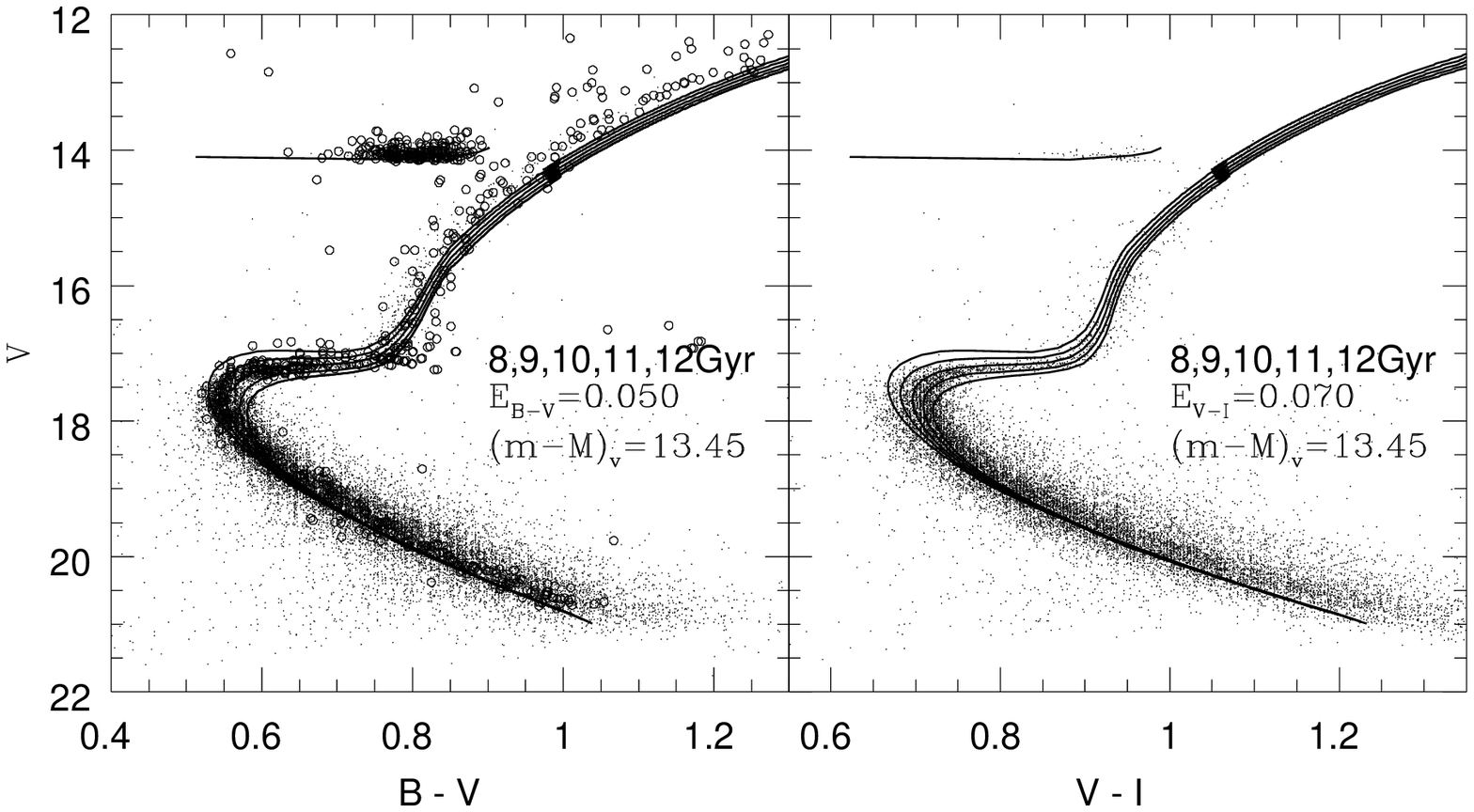,width=8in}}
\vskip-9.2cm
\caption{
In the left we plot the CMD of 47~Tuc (data of Kaluzny et al. 1998 plotted as small 
circles; that of Hesser et al.~1987 plotted as large circles). Overplotted are various 
isochrones of [Fe/H]=-0.7, [$\alpha$/Fe]=+0.4, Z=0.008 and atomic diffusion for 
different ages.
In the right we plot the V-I versus V diagram using data of 
Kaluzny et al. (1998).
}
\label{fig:CMD}
\end{figure}

We have calculated a new set of isochrones (initial metallicity [Fe/H]=$-$0.7)
with He and metals diffusion included as in Salaris et al. (2000), and 
using the same input physics and $\alpha$-enhancement as in SW98. 
Fig.~\ref{fig:CMD} shows the fit to the 47~Tuc CMD of both Kaluzny et al. (1998) 
and Hesser et al. (1987) using these isochrones. The derived age (determined 
from the TO {\it luminosity} once the distance is fixed) is 9-11~Gyr. 
We obtain similar results when fitting the V-I versus V diagram using data of 
Kaluzny et al. (1998), with the exception that a better fit is attained for the 
giant branch than in the B-V versus V diagram.

Fig.~\ref{fig:solut} shows the H$\gamma_{\sigma<130}$ measurement for the model 
spectra computed on the basis of these new isochrones; the CMD age as estimated 
from Fig.~\ref{fig:CMD} is still smaller by $\sim$2~Gyr than the minimum age 
allowed by the H$\gamma_{\sigma<130}$ measurement, but much more consistent with 
the H$\gamma$ age derived employing the same set of isochrones
than obtained before. The right panels of Fig.~\ref{fig:solut} 
show that use of these new models does not significantly change our metallicity 
estimate for the cluster.

%%%%%%%%%%%%%%%%%%%%%%%%%%%%%%%%%%%%%%%%%%%%%%%%%%%%%%%%%%%%%%%%%%%%%%%%%%%%%%%%%%%%%
%%%%%%%%%%%%%%%%%%%%%%%%%%%%%%%%%%%%%%%%%%%%%%%%%%%%%%%%%%%%%%%%%%%%%%%%%%%%%%%%%%%%%
\section{Conclusions}
%%%%%%%%%%%%%%%%%%%%%%%%%%%%%%%%%%%%%%%%%%%%%%%%%%%%%%%%%%%%%%%%%%%%%%%%%%%%%%%%%%%%%
%%%%%%%%%%%%%%%%%%%%%%%%%%%%%%%%%%%%%%%%%%%%%%%%%%%%%%%%%%%%%%%%%%%%%%%%%%%%%%%%%%%%%
We have discussed the origin of the discrepancy between the spectroscopic (based 
on the effective temperature of TO stars) and CMD (based on the luminosity of TO 
stars and an assumed distance scale) age estimate for 47~Tuc as raised by G99. 
For this purpose we have defined a new age indicator, H$\gamma_{\sigma<130}$,
particularly suitable for studying GCs and low velocity dispersion galaxies, which 
shows a superb power to break the age-metallicity degeneracy. H$\gamma_{\sigma<130}$ 
confirms the age discrepancy found by G99 for 47~Tuc. 
Emission fill-in of the Balmer lines appears to be an unlikely source of the
weak H$\gamma$ in 47~Tuc, since the ages derived from different Balmer lines
give discordant results if the hypothetical emission fill-in is corrected for, 
and since a composite of four other metal-rich Galactic
globular clusters shows the same weak H$\gamma$ phenomenon.
Thus the fact that other 
metal-rich GCs show very similar low Balmer values in comparison to the 
model predictions suggests a problem in the zero point of current stellar 
population models. It is worth noting that this zero point problem 
of the models with respect to the metal-rich GCs also works out for old
elliptical galaxies.

We therefore analyzed the possible causes of the problem by studying a number 
of input parameters of the evolutionary computations, and comparing the 
observed value of H$\gamma_{\sigma<130}$ with that derived from the synthesized 
integrated spectra. Neither the initial He content nor the HB have significant 
effects on the Balmer indices synthesized for 47~Tuc. However the inclusion of 
$\alpha$-elements enhancement and atomic diffusion in the evolutionary models 
provide spectroscopic ages which are much closer to the CMD derived 
ages. This occurrence constitutes a possible solution to the age-discrepancy 
between CMD and integrated spectrum ages of old metal-rich stellar populations. 

It is important to study if the age discrepancy is present in metal poor GCs.
For this purpose we need to expand the current stellar 
spectral libraries which feed the stellar population models (see V99), to extend
their predictions to lower metallicities.

\acknowledgments
We are grateful to the nonanonymous referee G. Worthey for very helpful
suggestions which helped to improve the final version of the paper.
We thank S. Covino and L. Jones for providing us with spectra of 47~Tuc and
M~32 respectively. We also thank A. Bressan for a useful discussion.
A.V. acknowledges the support of the PPARC rolling grant 'Extragalactic
Astronomy and Cosmology in Durham 1998-2002'. 

%%%%%%%%%%%%%%%%%%%%%%%%%%%%%%%%%%%%%%%%%%%%%%%%%%%%%%%%%%%%%%%%%%%%%%%%%%%%%%%%%%%%%
%%%%%%%%%%%%%%%%%%%%%%%%%%%%%%%%%%%%%%%%%%%%%%%%%%%%%%%%%%%%%%%%%%%%%%%%%%%%%%%%%%%%%

%%%%%%%%%%%%%%%%%%%%%%%%%%%%%%%%%%%%%%%%%%%%%%%%%%%%%%%%%%%%%%%%%%%%%%%%%%%%%%%%%%%%%
\onecolumn
%%%%%%%%%%%%%%%%%%%%%%%%%%%%%%%%%%%%%%%%%%%%%%%%%%%%%%%%%%%%%%%%%%%%%%%%%%%%%%%%%%%%%

\end{document}